# An Optical Readout TPC (O-TPC) for Studies in Nuclear Astrophysics With Gamma-Ray Beams at HIγS[*]

M. Gai[1,2], M.W. Ahmed[3], S.C. Stave[3], W.R. Zimmerman[1,2], A. Breskin[4],
B. Bromberger[5], R. Chechik[4], V. Dangendorf[5], Th. Delbar[6], R.H. France III[7],
S.S. Henshaw[3], T.J. Kading[1], P.P. Martel[3,8], J.E.R. McDonald[9,2], P.-N. Seo[1,3],
K. Tittelmeier[5], H.R. Weller[3], and A.H. Young[1].

1. LNS at Avery Point, University of Connecticut, Groton, CT 06340-6097, USA
2. Dept. of Physics, Yale University, New Haven, CT 06520-8124, USA
3. TUNL, Dept. of Physics, Duke University, Durham, NC 27708, USA
4. Dept. of Particle Physics, Weizmann Institute of Science, 76100 Rehovot, Israel
5. Physikalisch-Technische Bundesanstalt, 38116 Braunschweig, Germany
6. Université Catholique de Louvain, 1348 Louvain-la-Neuve, Belgium
7. Georgia College & State University, CBX 82, Milledgeville, GA 31061, USA
8. Dept. of Physics, University of Massachusetts, Amherst, MA 01003, USA
9. Dept. of Physics University of Hartford, West Hartford, CT 06117-1599, USA

## ABSTRACT

We report on the construction, tests, calibrations and commissioning of an Optical Readout Time Projection Chamber (O-TPC) detector operating with a $CO_2(80\%) + N_2(20\%)$ gas mixture at 100 and 150 Torr. It was designed to measure the cross sections of several key nuclear reactions involved in stellar evolution. In particular, a study of the rate of formation of oxygen and carbon during the process of helium burning will be performed by exposing the chamber gas to intense nearly mono-energetic gamma-ray beams at the High Intensity Gamma Source (HIγS) facility. The O-TPC has a sensitive target-drift volume of $30 \times 30 \times 21$ cm$^3$. Ionization electrons drift towards a double parallel-grid avalanche multiplier, yielding charge multiplication and light emission. Avalanche-induced photons from $N_2$ emission are collected, intensified and recorded with a Charge Coupled Device (CCD) camera, providing two-dimensional track images. The event's time projection (third coordinate) and the deposited energy are recorded by photomultipliers and by the TPC charge-signal, respectively. A dedicated VME-based data acquisition system and associated data analysis tools were developed to record and analyze these data.

The O-TPC has been tested and calibrated with 3.183 MeV alpha-particles emitted by a $^{148}$Gd source placed within its volume with a measured energy resolution of 3.0%. Tracks of alpha and $^{12}$C particles from the dissociation of $^{16}$O and of three alpha-particles from the dissociation of $^{12}$C have been measured during initial in-beam test experiments performed at the HIγS facility at Duke University. The full detection system and its performance are described and the results of the preliminary in-beam test experiments are reported.

_________________________________________________

[*] Supported by the Richard F. Goodman Yale-Weizmann Exchange Program, ACWIS, NY, and USDOE grant Numbers: DE-FG02-94ER40870 and DE-FG02-97ER41033.

# 1. Introduction

The development of the Optical Readout Time Projection Chamber (O-TPC) is motivated by the need to perform accurate measurements of cross sections of nuclear reactions that are essential for stellar evolution theory, and most importantly the measurement of cross sections relevant for stellar helium burning. We refer the reader to Ref. [1] for a thorough review of stellar evolution theory and the definition of the nomenclatures used in this field. The outcome of helium burning is the formation of the two elements: carbon and oxygen [1]. The ratio of carbon-to-oxygen at the end of helium burning has been identified three decades ago as one of the key open questions in Nuclear Astrophysics [1] and it remains so today. To solve this problem one must determine the p-wave [$S_{E1}(300)$] and d-wave [$S_{E2}(300)$] cross section S-factors, defined in Ref [1], of the $^{12}C(\alpha,\gamma)^{16}O$ reaction at the Gamow peak (300 keV) with an accuracy of approximately 10% or better [1].

Several new measurements of the $^{12}C(\alpha,\gamma)^{16}O$ reaction using gamma-ray detectors have been reported [2, 3] with center of mass energies in the vicinity of 1.0 MeV. However, the astrophysical cross section S-factors were determined with very low accuracies (±40-80%) and, most importantly, one cannot rule out a low value (close to zero) of the extrapolated E1 S-factor [4, 5]. These new experiments used some of the highest intensity alpha-particle beams (100 - 500 μA) with impressive luminosities of $10^{33}$ cm$^{-2}$sec$^{-1}$ [2] and $10^{31}$ cm$^{-2}$sec$^{-1}$ [3], and 4π arrays of HPGe and BaF$_2$ detectors, respectively, that provided large counting statistics. Yet the accuracies of the measured S-factors were limited by the quality of the measured angular distributions needed to separate the E1 and E2 components. A major disadvantage of measuring gamma-rays is the large background from neutrons emitted from the $^{13}C(\alpha,n)$ reaction, room background gamma-rays, cosmic rays and Compton scattering. Such a background is not expected in our proposed experiment with the O-TPC. In addition we will measure detailed angular distributions, thus obtaining accurate values of the E2/E1 ratio that are crucial for an accurate extrapolation to stellar energies (300 keV). We however note that the proposed experiment can only measure direct-capture to the ground state. The small (5-10%) contribution of cascade gamma-rays from the $^{12}C(\alpha,\gamma)^{16}O$ reaction will not be measured in this experiment.

The O-TPC concept [6] consists of recording optically, in three dimensions, track images formed within large gas volumes. The detection is based on capturing light emitted from avalanches induced by ionization electrons drifting into the detector's multiplying element. The three-dimensional images are reconstructed from projected two-dimensional track-images in an intensified Charged Coupled Device (CCD) camera and the third dimension from the projected time-structure recorded by photomultiplier tubes (PMTs) [6]. Information on optical readout of gaseous detectors, on photon emission from charge avalanches and on potential applications can be found elsewhere [7-10]. The rather low cost of the optical readout system and the simplicity of the data collection and analysis make such detectors ideally suited for the search of simple (e.g. two-body decay) rare events of known patterns in a low background environment. However, since its introduction over two decades ago, the O-TPC has had a limited use in Nuclear Research [11, 12]. The O-TPC concept on the other hand appears well suited for experiments at gamma-ray facilities such as the HIγS at the Triangle Universities Nuclear Lab (TUNL) [13,14] as used in the present study with intense gamma-ray beams [15].

The design of the current detector relies on extensive studies with a prototype O-TPC, in which an appropriate oxygen-rich target-gas ($CO_2$) mixed with a light emitting gas ($N_2$), was selected [16]. In this article we discuss the O-TPC detector, its optical readout elements along with the data-acquisition (DAQ) system and analysis concepts. The results of system characterization and calibrations with a $^{148}$Gd radioactive source and extensive tests with gamma-ray beams at the HIγS facility are also discussed.

## 2. The O-TPC Detector System
### 2.1 The TPC

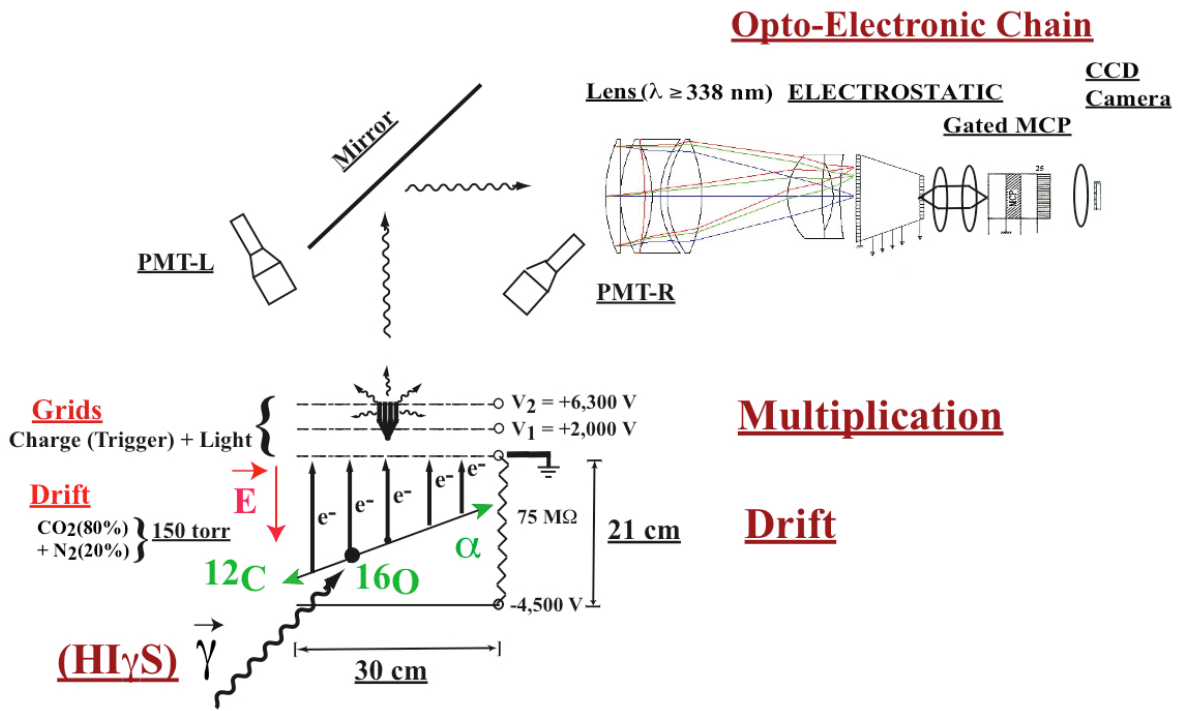

Fig. 1: A schematic diagram of the Optical Readout Time Projection Chamber (O-TPC) with an $^{16}$O target nucleus dissociated by the gamma-ray beam from HIγS. The gas mixture and pressure as well as the operating voltages are indicated. The drift electrons that provide the third (time projection) upward dimension are shown schematically.

A schematic diagram of the O-TPC detector with the optical readout chain including a CCD camera to record the track-images and the PMTs to record the time-projected coordinate is shown in Fig. 1. The four walls of the drift chamber region, with dimensions of 30x30 cm$^2$ and 21.0 cm height, are made out of double-sided G10 printed circuit boards (PCB) with 66 copper strips each 2.5 mm wide with 0.4 mm spacing, establishing a very homogenous drift field. The copper strips are interconnected through 1 MΩ resistors to form a high-voltage divider, as shown in Fig. 2. The gamma-ray beam (as well as the calibration source) enters and exits the drift volume via 15 mm diameter holes placed in the center of the front and back walls of the drift cage, as shown in Fig. 2. The 30x30 cm$^2$ square cathode and anode grids are placed 5 mm below and above the last copper strips, with 5 MΩ resistors connected to the adjacent copper strip. With the O-TPC operating at 150, and 100 Torr, a negative high voltage of -4,500 V and -3,500 V, respectively, is applied to the cathode placed at the bottom of the drift volume. The top anode grid is kept at ground potential yielding a current of ~60 μA and ~47 μA, respectively, through the high-voltage divider chain.

The grid electrodes are made from a stainless steel wire mesh, with 50 μm thick wires separated by 500 μm (~80% optical transmission) soldered to 33x33 cm$^2$ PCB G10 frames. The three grids (including the anode) are separated by 5 mm, forming a double parallel-grid charge multiplication and light emitting structure. Following our extensive light-emission studies [16] we chose the $CO_2$(80%) + $N_2$(20%) gas mixture (99.999% purity) in order to optimize the oxygen content and the emitted light yield and wavelength (primarily 338 nm). The detector is pumped down (1 mTorr) over a period of several months before use to reduce out-gassing and it is operated with gas flow (approximately 100 stcc/min) at 150 (and 100) Torr to optimize the track length of the gamma-ray induced charged particles; e.g. 34 mm and 7 mm for 1.8 MeV alpha-particles and 0.6 MeV $^{12}$C, respectively, at 150 Torr.

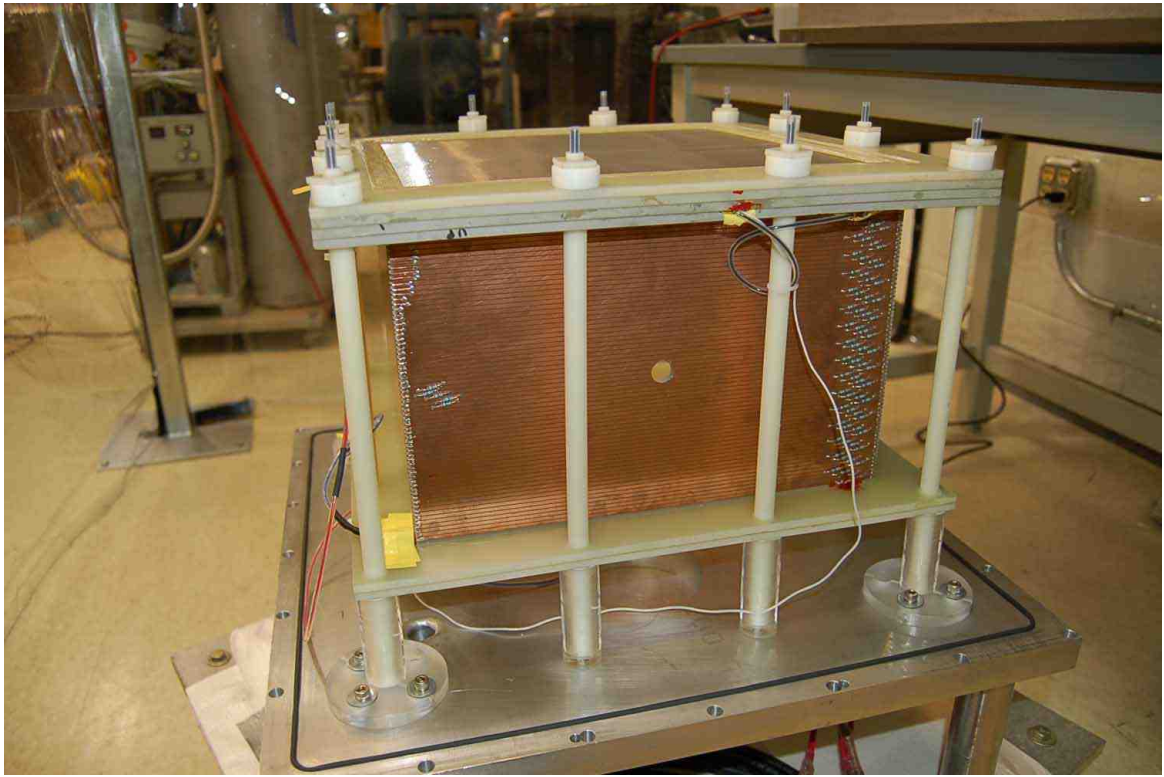

Fig. 2: Photograph of the O-TPC with its resistor chain along the field cage, beam entrance hole and the three parallel-grid multiplying structure (at the top).

## 2.2 The Opto-electronic Readout System

The opto-electronic system is comprised of PMTs and an intensified CCD camera system. During the current tests, two 75 mm diameter PMTs with quartz windows view the TPC from a distance of approximately 85 cm through a mirror placed above the 2.5 cm thick (40 cm diameter) quartz window of the TPC. At a later stage four Hamamatsu R10133 PMTs were placed closer (approximately 60 cm) to the last grid and directly above the quartz window at the location of the two PMTs shown in Fig. 1. The PMTs operate with a high voltage of -1,600 V. They provide a measurement of the time projection along the direction of the drifting electrons (drift axis). The light emitted from a tilted track arrives at the PMT with a typical spread of a few micro-seconds due to the (slow: approximately one cm/µs) drift velocity of the electrons in the gas. The digitized signals of the PMTs are summed for measuring the pulse-shape along the drift-coordinate. As we discuss later the pulse shape of the PMT signal provides a measurement of the out of (horizontal) plane angle.

The 2D track images in the (horizontal) plane perpendicular to the drift axis are recorded with the opto-electronic chain shown in Fig. 3. As we discuss later the image of the track provides a measurement of the in-plane angle of the track. The opto-electronic chain consists of a large lens, an electrostatic demagnifier, two commercial (35 mm diameter) lenses connected front-to-front, a gated Micro Channel Plate (MCP), and a CCD camera. The custom-designed lens [17] shown in Fig. 3 with diameter D = 142 mm and effective F-number (=f/D) f/1.4 was produced by Optimax [18]. The lens is placed at a distance of 635 mm from the top grid, with a 370 mm diameter field of view. The image of 100 mm diameter (demagnification = 3.7) is 18 mm behind the back end of the lens. During some of the current studies a considerably smaller-diameter (60 mm) lens was used, with a 10-fold lower light collection efficiency.

The image of the lens is projected on the photocathode of a Hamamatsu V4440U demagnifier previously used by the Université Catholique de Louvain (UCL) group in the CERN-CHORUS experiment [19]. We refer the reader to this reference for details. The Quantum Efficiency (QE) of its photocathode, shown in Fig. 4, was measured at the Physikalisch-Technische Bundesanstalt (PTB) at Braunschweig, Germany. A large QE (20%) was measured at 338 nm, the wavelength of the majority of the light emitted by nitrogen [16]. Note that 3.18 MeV alpha-particle tracks viewed by the 142 mm diameter lens yield approximately 1000 photo-electrons in the photo-cathode of the Hamamatsu demagnifier. The spatial resolution of the demagnifier was measured using a calibration mask illuminated by an electro-luminescent screen. The measured Contrast Transfer Function (CTF) is shown in Fig. 5.

The 25 mm diameter phosphor screen of the demagnifier is viewed by two 35 mm diameter commercial lenses with F-number f/2.8 connected front-to-front, as shown in Fig. 3. This permits placing them very close to the screen, with good light collection efficiency. The two 50 mm focal length lenses, when placed 50 mm from the phosphor screen of the demagnifier, project an image on the photocathode of the MCP image intensifier placed 50 mm behind the second lens. The long (50 µsec) decay time of the screen permits a delayed arrival of a trigger signal to record the track image with the CCD camera shown in Fig. 3, that is placed behind the triggered MCP viewing the exit phosphor screen (P11) of the demagnifier.

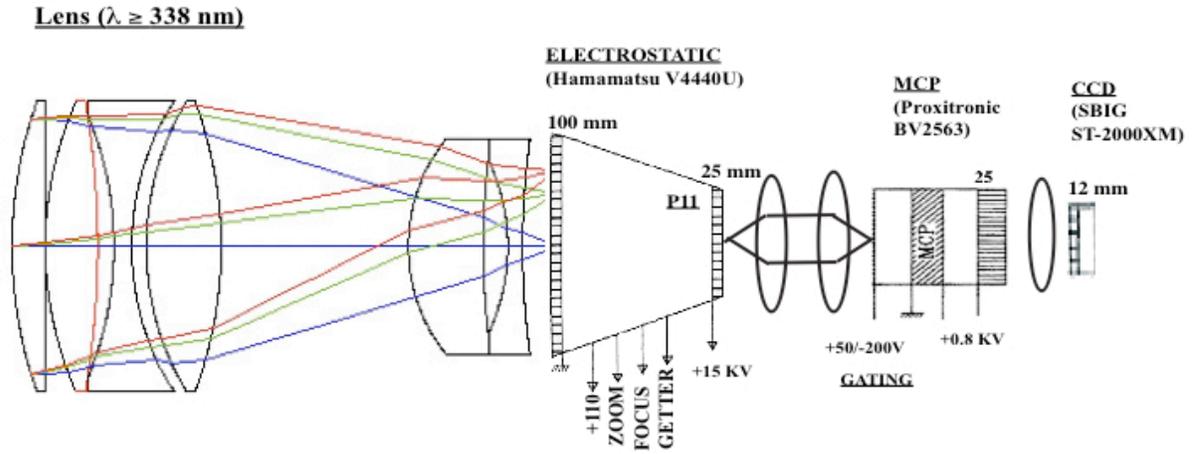

Fig. 3: A schematic diagram of the UCL-PTB-TUNL opto-electronic chain used in the O-TPC.

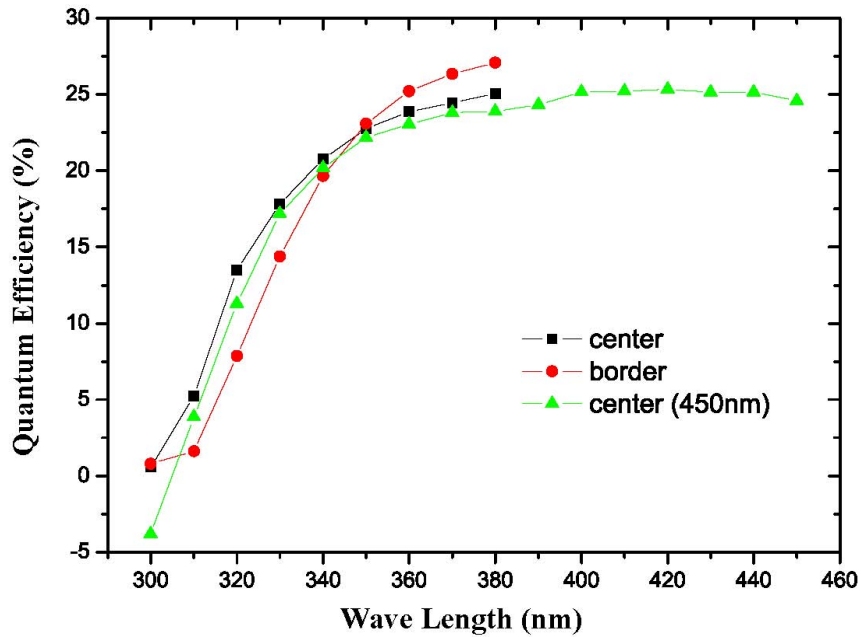

Fig. 4: The measured Quantum Efficiency (QE) of the 100 mm diameter photo-cathode of the entrance window of the Hamamtsu V4440U demagnifier. For the 300 – 380 nm scan two irradiation spots were used separated by 35 mm; one in the center (black squares) and a second one in the periphery (red circles). For the 300 – 450 nm scan only one irradiation spot was used in the center (green triangles).

The image intensifier is a gated Proxitronics MCP model BV2563. The gate of the MCP is generated by the grid (charge) signal of the TPC (with a deposited-energy threshold of 1.0 MeV), in coincidence with the two PMTs. When the gate is opened by the NIM trigger signal, a +50 V bias voltage placed on the photocathode of the MCP is changed to -200 V, allowing the photo-electrons to leave the photo-cathode of the MCP. This gate allows for effectively discriminating against low energy (~20 keV) gamma-ray induced electrons produced in the gas. It reduces the background trigger rate down to 0.02 Hz. A cooled (-5º C) CCD camera, SBIG model ST-2000XM [20], shown in Fig. 3, is used to capture the track image present at the back phosphor screen of the gated MCP.

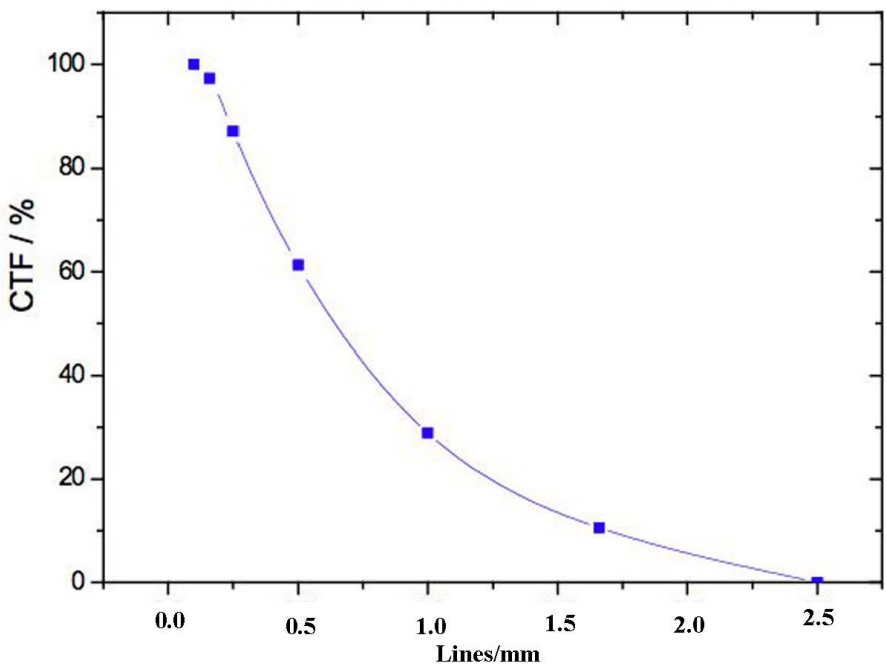

Fig. 5: The measured Contrast Transfer Function (CTF) as a function of line frequency (lines per mm) of the Hamamatsu demagnifier.

## 3. The VME Data Acquisition and Camera Readout System

In this section we discuss the logic circuitry used to collect the data and capture the digital image of the track. At the beginning of each run a dark image exposure is taken in order to subtract it from captured images and thus remove "hot pixels" in the CCD image. Following dark-image storage, the shutter of the camera is kept open until the appearance of a valid event. When the energy deposited in the TPC is larger than 1.0 MeV the gated MCP is triggered yielding a track image on its screen that is recorded by the CCD camera. The same signal triggers the VME data acquisition system where the PMT signals are digitized with a 100 MHz SIS3301 Struck flash analog-to-digital converter (FADC). In addition the analog signals from the PMTs as well as the charge signal from the grid are shaped with an ORTEC 673 spectroscopy amplifier (with a 6 μsec shaping time) and digitally converted with an ADC. The shaping of the PMT analog signal resulted in a PMT histogram with better resolution than the one obtained by digital integration of the output of the FADC. Scalers are used to register counts from gamma-ray beam monitors and record the number of events. For each event the content of the digitized FADC, analog ADC, CCD camera and scalers are downloaded in the form of one event recorded by a DAQ computer running CODA software [24] on a Linux operating system. During this download process a veto prohibits triggering of new events. Due to the long download time (typically a few seconds) of the currently used CCD camera, the maximum useful event rate that can be handled by our DAQ system is of the order of 0.2 Hz with a live-time on the order of 60%. The event rate of the DAQ system in the current study with $CO_2 + N_2$ gas mixture is dominated by the approximately 1.9 MeV background protons (for $E\gamma$ = 9.55 MeV) from the $^{14}N(\gamma,p)$ reactions that deposit more than 1.0 MeV in the detector (i.e. above the trigger threshold). For beam intensities above $10^8$ γ/sec a (background) trigger event rate of a few Hz is anticipated, and the currently used SBIG-CCD camera is too slow. We installed a considerably faster camera, Hamamatsu ORCA-R2, capable of recording 30 images per second and it will be used in future studies.

## 4. The Coordinate System

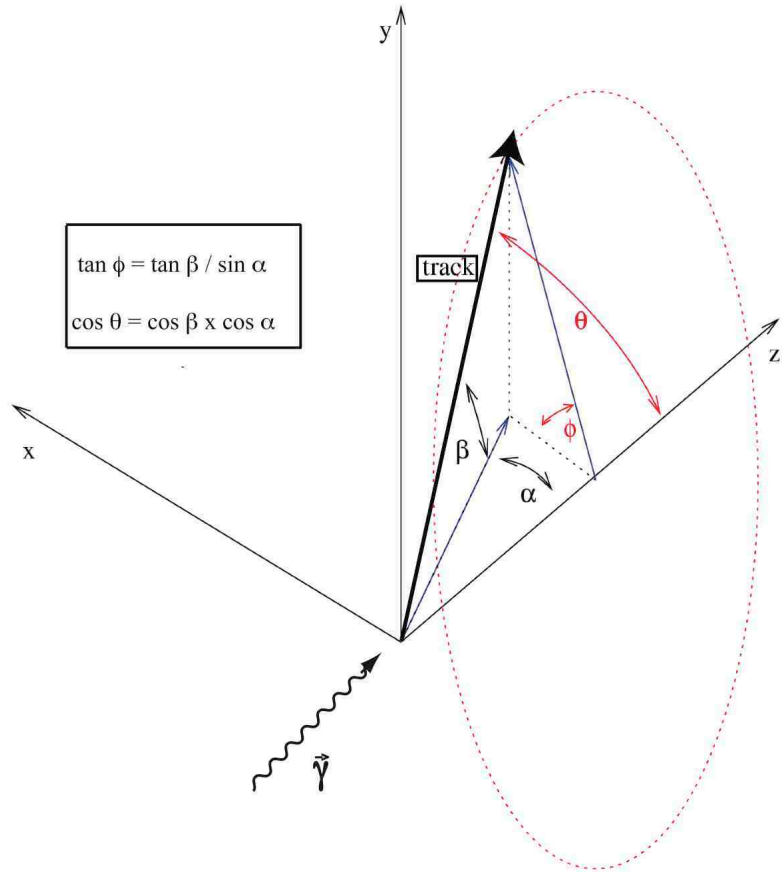

Fig. 6: The horizontal in-plane angle ($\alpha$) measured by the projected track (i.e. CCD image) and the vertical out-of-plane angle ($\beta$) measured by the time projection (i.e. PMT signal). They are transformed into the scattering angle ($\theta$) and the azimuthal angle ($\phi$) of the spherical coordinate system used in scattering theory using the equations shown in the box.

The coordinate system of the O-TPC is shown in Fig. 6. The origin is at the center of the drift volume, the z-axis is along the beam direction with its positive direction pointing downstream of the beam, the x-axis points toward beam left and the y-axis points upward, forming the conventional right handed (x, y and z) Cartesian coordinate system. As discussed below the projected image of the track, recorded by the CCD camera, defines the horizontal in-plane angle ($\alpha$). The time projection derived from the pulse shape of the PMT's light-signal defines the vertical out-of-plane angle ($\beta$). The transformation of the angles ($\alpha,\beta$) to the ($\theta,\phi$) scattering and azimuthal angles (of the spherical coordinate system) used in scattering theory is given by:
$\tan \phi = \tan \beta / \sin \alpha$ and $\cos \theta = \cos \beta \times \cos \alpha$; as shown in Fig. 6.

## 5. Performance of the O-TPC
## 5.1 Source Tests

The electron multiplication curves as well as the light detected by the PMTs (placed at a distance of approximately 85 cm from the TPC cahmber) are shown in Fig. 7. The measurements were done at a pressure of 150 Torr by varying the voltages of the first and the second grids with respect to ground as shown in Fig. 1. Gains of ~$10^4$ were reached with 3.18 MeV alpha-particles. The gain was determined by measuring the total charge deposited on the grid divided by the original ionization charge produced by the alpha-particles. Based on these results we opted to use the voltages indicated in Fig. 1, with the anode at ground potential, the first grid at $V_1$ = +2,000 V ($\Delta V_1$ = +2,000 V over a 5 mm gap) and the next grid at $V_2$ = +6,300 V ($\Delta V_2$ = +4,300 V over a 5mm gap with the field (E) divided by the pressure (P); i.e. the reduced field E/P = 57.3 V/cmTorr). The cathode voltage was kept at -4,500 V, resulting in a reduced drift field of E/P = 1.25 V/cmTorr.

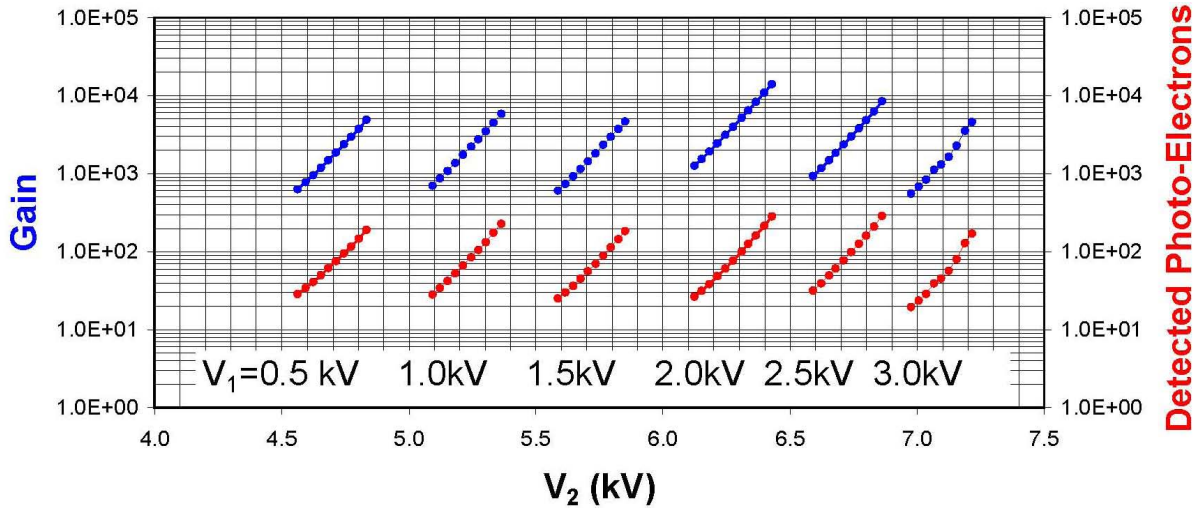

Fig. 7: Gain curves (upper blue curves) measured in the TPC with a 10 nCi $^{148}$Gd source at 150 Torr $CO_2$(80%) + $N_2$(20%) and the number of photo-electrons (lower red curves) detected in the PMTs, as a function of the voltage of the top grid ($V_2$) for various values of the voltage of the grid below it ($V_1$) as shown in Fig. 1.

The last grid is connected via a large (20 nF) capacitor and a high voltage filter box to a Canberra 2003B preamplifier. When needed, these signals are attenuated using a capacitor connected to ground. The preamplified signals are shaped by a standard spectroscopy amplifier with 6 µs shaping time and processed by a VME based ADC. A collimated 10 nCi $^{148}$Gd (0.3 Hz) source is inserted into the TPC using a long Lucite rod. The spectra of the charge produced by the 3.183 MeV alpha-particles and of the light collected in the PMTs are shown in Fig. 8. Energy resolution of 2.5-3.0% (FWHM) is observed over a long period of time with the source placed at one location within the drift volume. Note the low background in the histograms shown in Fig. 8. The poorer PMT resolution observed here is primarily the result of the small number of photo-electrons (see Fig. 8).

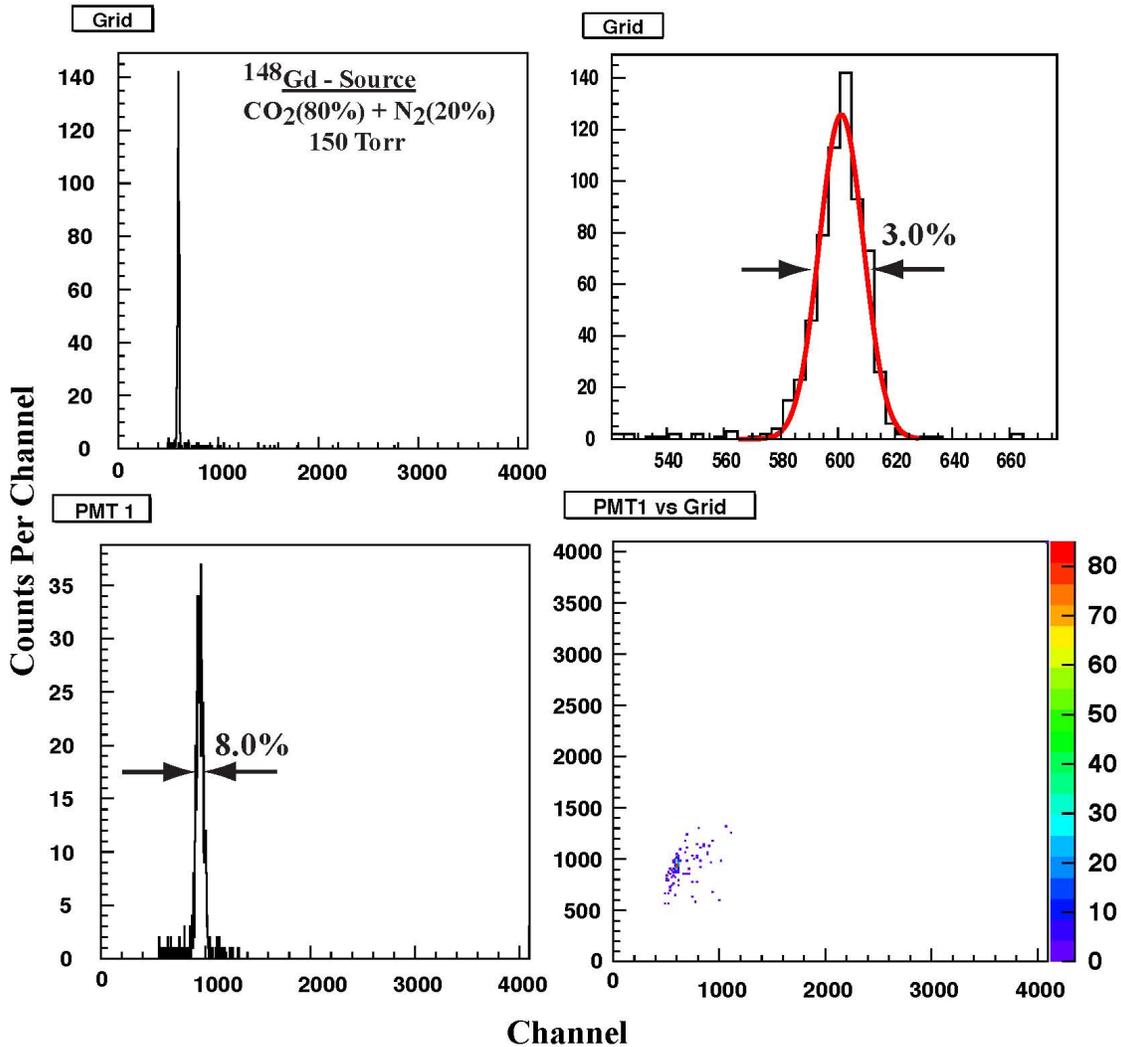

Fig. 8: TPC Spectra of alpha-particles from a $^{148}$Gd source measured with the $CO_2$(80%) + $N_2$(20%) gas mixture at 150 torr. We show the energy grid signal (top left), the light PMT1 signals (bottom left) and light-vs-charge correlation (bottom right). The region of interest of the charge signal with a Gaussian fit (red curve) is shown (top right). Note the low background in the histograms shown here. The poor(er) PMT resolution is due to low number of photo-electron as discussed in the text.

The pulse-height variations along the beam axis, shown in Fig. 9, were measured by displacing the source. A significant variation of the signal was measured along the z-axis and for beam-left (positive x) versus beam-right (negative x); note the abscissa calibration (for these data): 1.1 pixels per mm. These variations were found to depend on the voltages applied to the grids. Hence we conclude that they are due to the mutual electrostatic attraction of the grids that reduces the distance between the grids; e.g. we estimate that in our geometry a 30 μm gap variation would affect the electric field causing up to ~20% variation in charge multiplication. The pulse height corrections of the photo-dissociation data were deduced from measuring the pulse height variation of the measured events. The measured pulse height signals were corrected with graphs similar to the one shown in Fig 9 to yield the corrected total-energy histograms discussed below.

The electron drift velocity of 1.1 cm/μs (at 100 Torr) was determined by measuring the line shape of the PMT signal from a well collimated source and comparing it with the predicted line shape based on energy loss dE/dx calculations using SRIM [21]. The measured drift velocity agrees quite well with Magboltz calculations of 1.14 cm/μs [22] and this value was used in the fits. The same Magboltz calculations yield transverse straggling of the drift electrons over 10 cm which is smaller than 1 mm. In addition 2 mm lateral straggling (only) at the end of the track is predicted by SRIM for the stopped alpha-particles [21]. The range of delta electrons in the gas used in this study is estimated using previous data [23] scaled to the current gas density (at 100 Torr) and is found to be 0.8 mm.

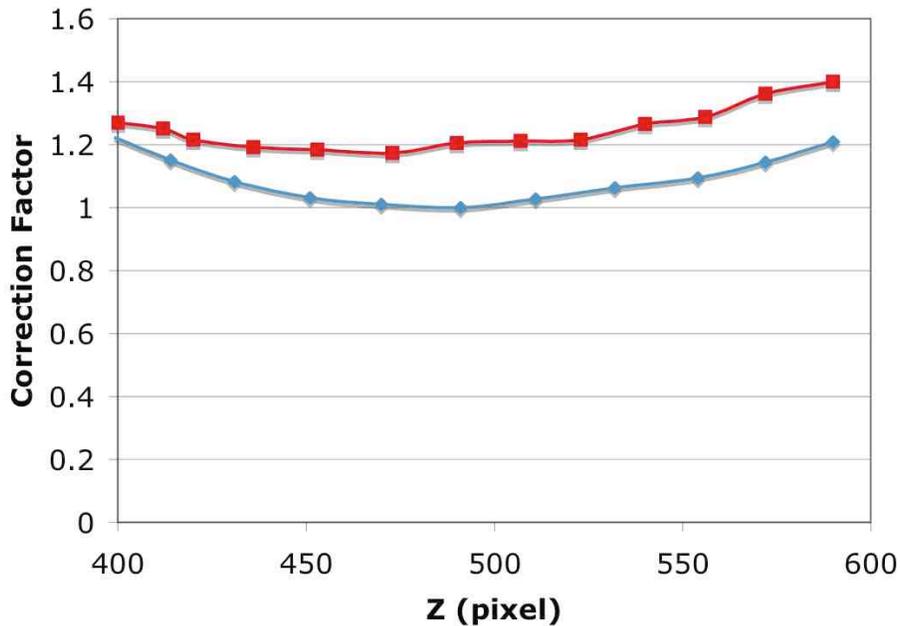

Fig. 9: The (z) position dependent correction of the grid charge signal for 3.183 MeV alpha-particles emitted at right angles on the beam-left (blue diamond, α = 90º) and the beam-right (red square, α = 270. The calibration (for these data) is 1.1 pixels/mm. The statistical error bars are smaller than the size of the symbol used.

## 5.2 In-Beam Tests

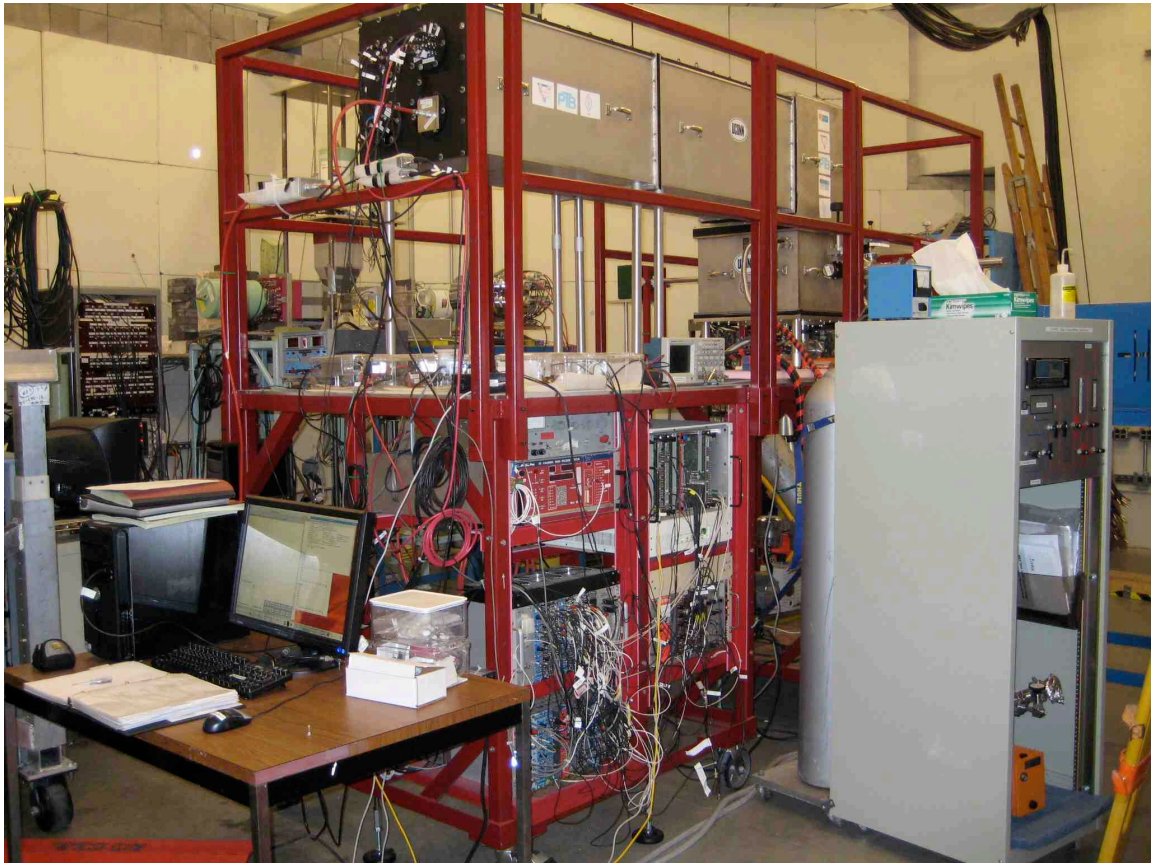

Fig. 10: A photograph (back view) of the O-TPC setup placed on a movable platform in the beam line at HIγS. The beam enters the TPC from the front and the source is inserted from the back. Shown are the gas handling system (back right), TPC, mirror box, optical readout box, CAEN high-voltage power supply, electronics and VME DAQ system

Several in-beam engineering runs were performed with the O-TPC located at the HIγS facility as shown in Fig. 10. Linearly polarized gamma-ray beams ranging between 9.5 and 11.1 MeV with average intensity of $1.0 - 4.5 \times 10^7$ γ/s on target were collimated by a 150 mm long, 12 mm diameter, lead collimator before entering the O-TPC. Two lead plugs 4 mm long having a diameter of 4 and 8 mm were placed in front and behind the detector, respectively, and the shadow of the lead plugs were captured by a gamma camera [24] placed behind the detector. This procedure allowed us to align the O-TPC detector with respect to the gamma-beam with an accuracy of 0.3 mm. A 200 μm thick Kapton entrance window was placed approximately 50 cm before the O-TPC with a strong magnetic field (500 Gauss) to deflect the electrons produced at the Kapton window. The interaction point of the 9.771 MeV gamma-ray beam with oxygen nuclei to produce $\alpha + {}^{12}C$ tracks was measured in this analysis by determining the end of the $^{12}C$ track (and knowledge of the length of the $^{12}C$ track) as shown in Fig. 11. The FWHM of the extracted interaction points is equal to the beam diameter of 12 mm and the tail is due to the fluctuations in the measured edge of the $^{12}$C track. The centroid of the distribution of interaction points is at X = +2 mm indicating a +2 mm misalignment of the camera (for these data). The misalignment of the camera (for these data) is corrected by shifting the beam position on the track-image. This shift (+2 mm) is negligible as compared to the 12 mm width of the beam spread.

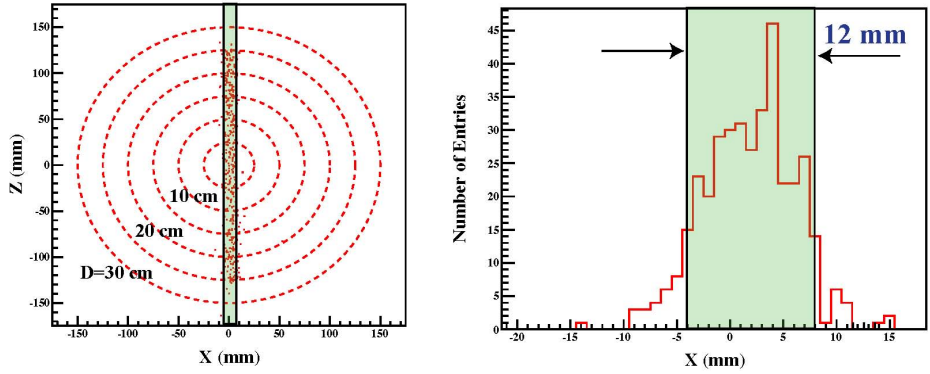

Fig. 11: Interaction points of $\alpha$ + $^{12}$C events from 9.771 MeV gamma-ray beams collimated with 150 mm long, 12 mm diameter, lead collimator. The beam spread is indicated in green shaded area.

Typical $\alpha$ + $^{12}$C events from the dissociation of $^{16}$O are shown in Fig. 12. Note that an approximately 9.5 MeV gamma-ray beam yields an approximately 1.8 MeV alpha-particle and an approximately 0.6 MeV $^{12}$C from the dissociation of $^{16}$O. A "three alpha-particle event" from the dissociation of $^{12}$C is shown in Fig. 13. The measured line shape of the PMT signal allowed us to reconstruct the track geometry shown in Fig. 13. All events were measured at a pressure of 100 Torr. Note that the track width is approximately 6.5 mm as shown in Fig. 12. The 63 mm track length of the horizontal (flat) $\alpha$ + $^{12}$C track shown in Fig. 12, is as expected from the range of the outgoing alpha-particle and $^{12}$C at 100 Torr [21]. Such $\alpha$ + $^{12}$C tracks are recorded by the CCD camera with close to 1000 photo-electrons on average, and they allow us to measure the track in-plane angle ($\alpha$) with an accuracy better than 3º. The out-of-plane angle ($\beta$) is measured from the line shape of the PMT signal (time projection signal). The summed PMT data shown in Fig. 12 (bottom right) from a different (tilted) track were fitted with the same line shape calculated from the dE/dx of the outgoing alpha-particle and $^{12}$C [21] folded with the track width and other relevant resolutions. Since we measured the drift velocity the fit shown in Fig. 12 has essentially one free parameter: the out of plane angle ($\beta$) which is extracted with an accuracy better than 3º. The scattering angle ($\theta$) and the azimuthal angle ($\phi$) are calculated for each event using the measured $\alpha$ and $\beta$ angles as discussed in section 4. The angular distributions are measured in 18 bins of $\Delta\theta = 10º$ each, considerably larger than the accuracy obtained from folding the in plane angles ($\alpha$) and the out of plane angle ($\beta$) to derive the scattering angle ($\theta$).

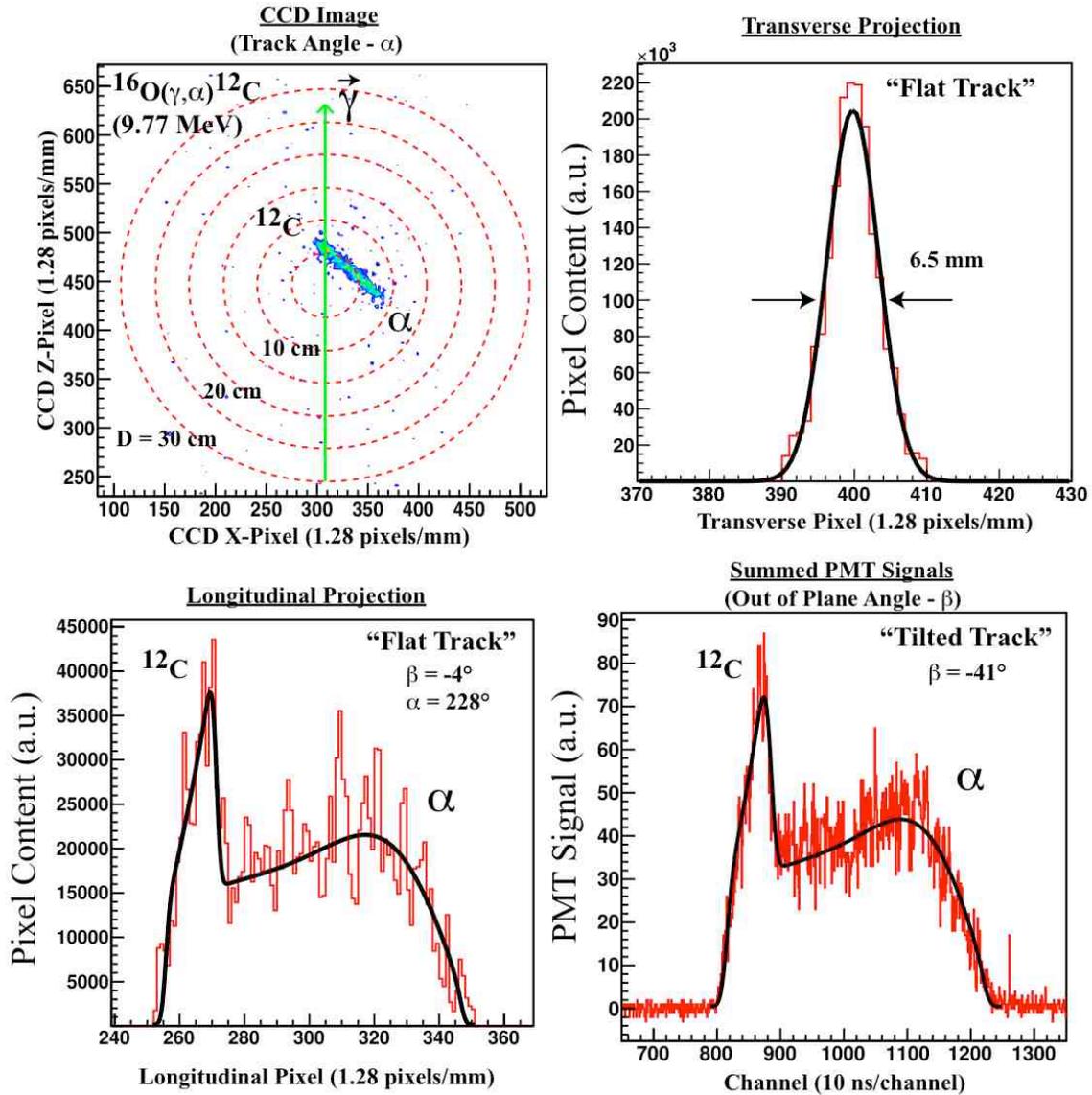

Fig. 12: A typical 63 mm long α + $^{12}$C track recorded at 100 Torr by the CCD camera (top left) from the dissociation of $^{16}$O by 9.77 MeV gamma-rays. The transverse (top right) and longitudinal (bottom left) projections of this nearly horizontal (β ≈ 0°) track are shown together with the fitted line shapes (Gassian and dE/dx, respectively). The PMT time projected pulse-shape (bottom right) of a different tilted track (β = −41°) is also shown together with the fitted dE/dx line shape (in black). The extracted in-plane angle (α) and out of plane angle (β) are indicated for the two different events shown in the bottom panels. The large light variations observed in the longitudinal projection arise from the fiber structure of the window in the back of the MCP.

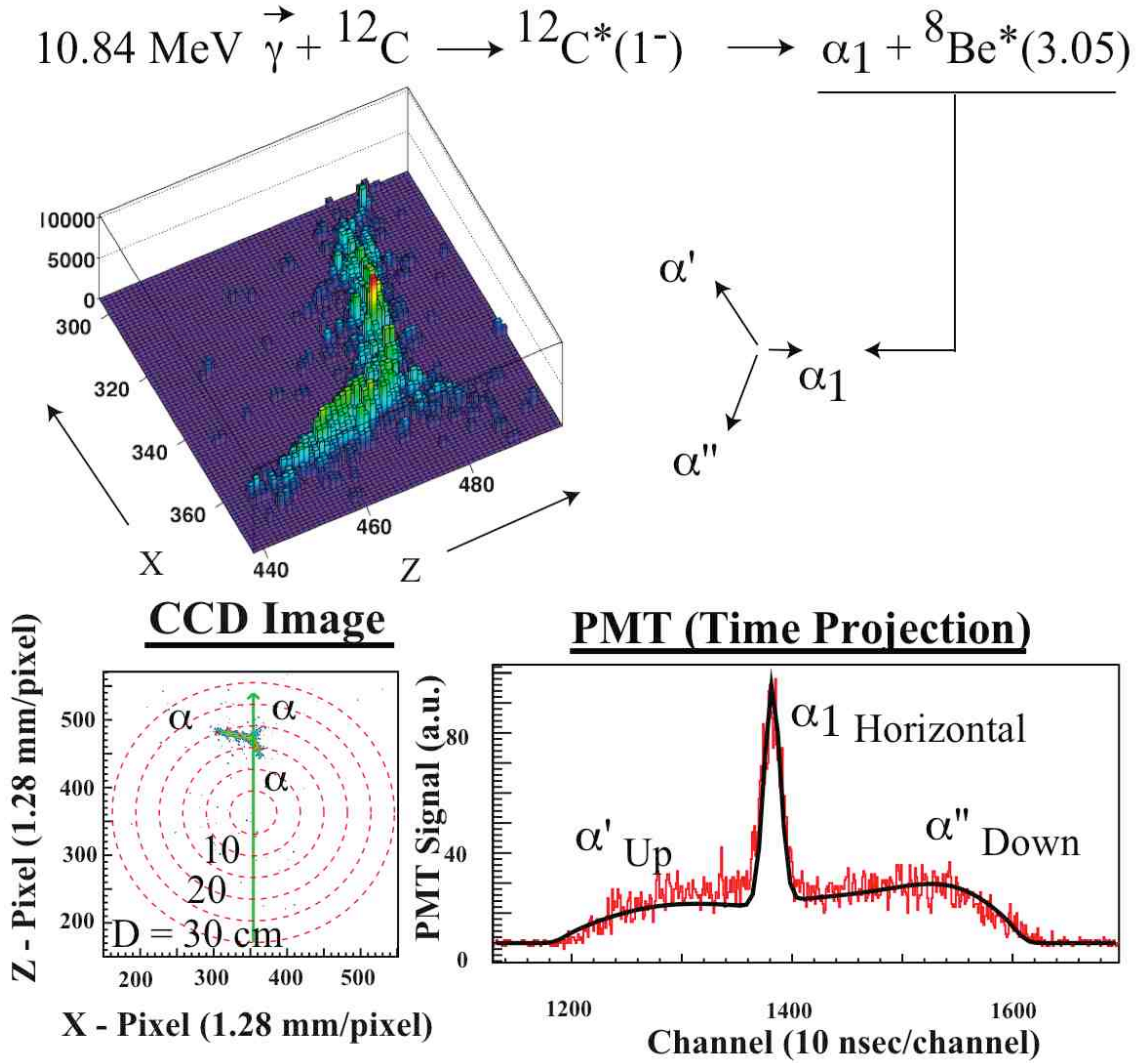

Fig. 13: Three alpha-particle event from the dissociation of $^{12}$C measured at 100 Torr. The projected 2D track and time of this event are shown in the lower part of the figure together with the fitted line shape of the light detected from the emerging three alpha-particles. The geometry of the reconstructed event is shown schematically with $\alpha_1$ emitted horizontally and the $^{8}$Be decay products, $\alpha'$ and $\alpha''$, emitted upward and downward, respectively.

The TPC charge signals are corrected for each event using graphs similar to the one shown in Fig. 9. This allows us to collect spectra for the total energy deposited in the dissociation of $^{16}$O (Q = 7.16195 MeV) and $^{18}$O (Q = 6.2270 MeV) as shown in Fig. 14. The $^{16}$O peak also includes events from the dissociation of $^{12}$C into three alpha-particles (Q = 7.27474 MeV, Q($\alpha + ^{8}$Be) = 7.3666 MeV). These spectra exhibit the FWHM of the gamma-ray beam which is approximately 310 keV. We note that the detector resolution (approximately 90 keV) is considerably better than the beam energy spread. Hence the data shown in Fig. 14 can be analyzed in smaller bins of approximately 100 keV each, and the cross section can be measured with approximately 100 keV resolution.

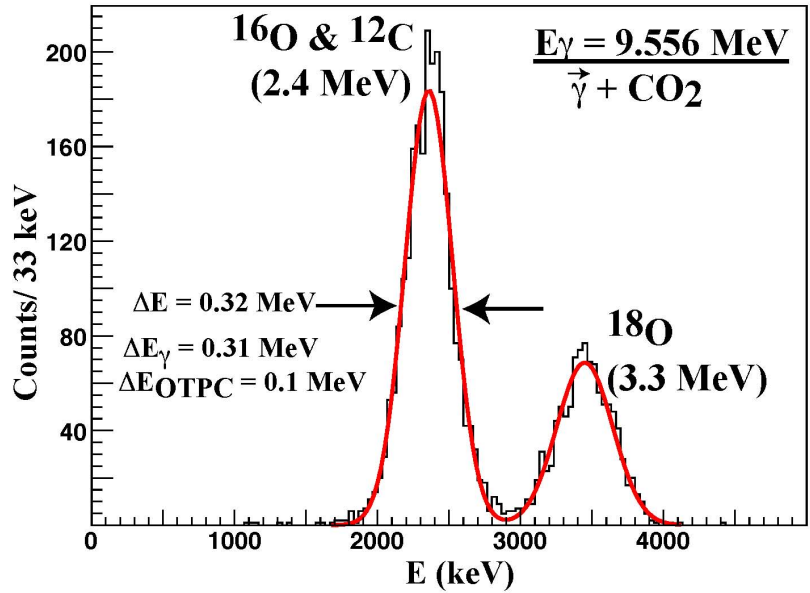

Fig. 14: Typical measured total energy (charge) spectrum from the dissociation of $CO_2$. The measured FWHM is dominated by the beam spread (FWHM ≈ 310 keV) as shown for the fitted Gaussian line shapes.

The current in-beam tests were performed with linearly polarized gamma-ray beams with intensities of up to $4.5 \times 10^7$ γ/s on target. A 3-fold larger intensity ($1.3 \times 10^8$ γ/s) of circularly polarized gamma-ray beam was also used for testing the O-TPC. The larger intensity of circularly polarized gamma-ray beam requires the faster Hamamatsu camera that was already installed in our setup to replace the SBIG camera that was used during these tests. The trigger rate is mostly due to protons from $^{14}N(\gamma,p)$ reactions (from the $CO_2 + N_2$ gas mixture). The background protons (1.9 MeV at Eγ = 9.55 MeV) have a range which is considerably larger than the active volume of the TPC, and therefore they deposit only a fraction of their energy in the O-TPC. Hence, by lowering the pressure to 100 Torr we removed most of the proton events that deposit less than 1.0 MeV in the gas and are below the trigger threshold.

Linearly polarized photons are well suited for measurements with the O-TPC since the outgoing particles for the reaction we are studying are emitted mainly in the horizontal plane, and the multiplied ionization charge is spread over the few cm long projected track. Circularly polarized gamma-rays, in contrast, do not yield a preferred plane of emission for the outgoing particles, thus outgoing particles can be emitted straight upward toward the grids. Such upward-traveling particles create large local electron densities (approximately $10^8$ electrons) - close to the Raether limit [25] which leads to discharges. Hence such upward-going events cannot be measured in the O-TPC detector. This reduces the usable solid angle to values smaller than 4π, and the measured data must be corrected for this deficiency.

## 6. Conclusions and Outlook

An O-TPC has been constructed, calibrated and tested with (3.18 MeV) alpha-particles from a $^{148}$Gd source and with intense (approximately 9.5 MeV) gamma-ray beams from the HIγS facility at Duke University incident on $^{16}$O and resulting in approximately 1.8 MeV alpha-particles and 0.6 MeV $^{12}$C. The O-TPC measures the total energy deposited with high resolution (e.g. 2.5-3.0% FWHM for 3.183 MeV alpha-particles). For each event we measure the scattering and azimuthal angles with an accuracy of ±3º, considerably smaller than the design goal bin size of 10º in the measured angular distributions. Complete angular distributions will be measured in 100 keV intervals, corresponding to the detector resolution and considerably better than the beam resolution (300 keV). The measured dE/dx along the track (together with the track length) and the line shape of the light signal allows for identifying $^{12}$C and $^{16}$O events with an estimated accuracy of ±10% or better in separating these events. Thus a complete kinematical characterization of each event is achieved with very high efficiency. The O-TPC detector measures all necessary kinematical variables over a large fraction of 4π, but the missing data (from upward moving tracks) need to be corrected. Such correction(s) will still allow us to measure the total cross section with the design goal accuracy of ±10%.

The O-TPC is indeed a prefect match for pursuing research with gamma-ray beams at the HIγS facility. It is ideally suited for low counting rate events with a simple (two body) structure. It is particularly useful for measuring angular distributions and cross sections of reactions that are essential for stellar evolution theory. We have commenced a program to measure the cross section of the $^{12}$C(α,γ)$^{16}$O reaction [1] by measuring the time reversed photo-dissociation of $^{16}$O. An additional experiment to measure the cross section of p(n,γ)d is in progress. The production rate of deuterium during Big Bang Nucleosynthesis is currently ill determined due to a substantial uncertainty in the measured values of the cross section. For this measurement research into the choice of gas is needed as we previously performed [16] for the case of $^{16}$O with a smaller prototype Optical Readout TPC.

## 7. Acknowledgements

The authors thank the Instrumentation Divisions at Brookhaven and at CERN for preparing the double-sided printed circuit boards. We thank Mr. Moshe Klin of the Weizmann Institute for preparing the grids and the gas handling system. We thank Mr. Guy van Beek of the Université Libre de Bruxelles and Mr. Bernard de Callatay of the Université Catholique de Louvain for their help in transferring the opto-electronic chain used by the CERN-CHORUS experiment. The technical staff at CERN, BNL, Weizmann, PTB, UCL and TUNL are acknowledged for numerous suggestions and assistance during this project. Dr. Y. Wu and the HIγS accelerator staff are acknowledged for delivering the high quality gamma-ray beams and Mr. M. Amamian is thanked for his help in aligning the O-TPC detector. A. Breskin is the W.P. Reuther Professor of Research in The Peaceful Use of Atomic Energy.

# 8. References

[1]  W.A. Fowler, Rev. Mod. Phys. **56**(1984)149.
[2]  M. Assuncao *et al.*; Phys. Rev. **C73**(2006)055801.
[3]  R. Plag et al.; Nucl. Phys. A758(2005)415c.
[4]  L. Gialanella *et al.*; Eur. Phys. J. A11(2001)357.
[5]  G.M. Hale; Nucl. Phys. A621(1997)177c.
[6]  P. Fonte, A. Breskin, G. Charpak, W. Dominik and F. Sauli; Nucl. Instr. Meth. **A283**(1989)658.
[7]  A. Breskin; Nucl. Phys. **A498**(1989)457C, and references therein.
[8]  G. Charpak, W. Dominik, J.P. Fabre, J. Gandaen, V. Peskov, F. Sauli, M. Suzuki, A. Breskin, R. Chechik and D. Sauvage; IEEE Trans. Nucl. Sci, **NS-35**(1988) 483.
[9]  D. Sauvage, A. Breskin and R. Chechik; Nucl. Instr. Meth.; **A275**(1989)351.
[10] L.M.S. Margato, F.A.F. Fraga, S.T.G. Fetal, M.M.F.R. Fraga, E.F.S. Balau, A. Blanco, R. Ferreira Marques, A.J.P.L Policarpo; Nucl. Instr. Meth. **A535**(2004) 231.
[11] U. Titt, A. Breskin, R. Chechik, V. Dangendorf, H. Schmidt-Bocking, H. Schuhmacher; Nucl. Instr. Meth. **A416**(1998)85.
[12] M. Cwiok, W. Dominik, Z. Janas, A. Korgul, K. Miernik, M. Pfützner, M. Sawicka, and A. Wasilewski; IEEE Trans. Nucl. Science, **52,#6**(2005)2895.
[13] V. N. Litvinenko, B. Burnham, M. Emamian, N. Hower, J. M. J. Madey, P. Morcombe, P. G. O'Shea, S. H. Park, R. Sachtschale, K. D. Straub, G. Swift, P. Wang, Y. Wu, R. S. Canon, C. R. Howell, N. R. Roberson, E. C. Schreiber, M. Spraker, W. Tornow, H. R. Weller, I. V. Pinayev, N. G. Gavrilov, M. G. Fedotov, G. N. Kulipanov, G. Y. Kurkin, S. F. Mikhailov, V. M. Popik, A. N. Skrinsky, N. A. Vinokurov, B. E. Norum, A. Lumpkin and B. Yang; Phys. Rev. Lett. **78**(1997)4569.
[14] H.R. Weller, M.W. Ahmed, H. Gao, W. Tornow, Y.K. Wu, M. Gai, and R. Miskimen; Prog. Part. Nucl. Phys. **62**(2009)257.
[15] M. Gai, A. Breskin, R. Chechik, V. Dangendorf and H.R. Weller; APH A, Heavy Ion Physics, **25**(2006)461; nucl-ex/0504003.
[16] L. Weissman, M. Gai, A. Breskin, R. Chechik, V. Dangendorf, K. Tittelmeier and H.R. Weller; Jour. Inst. **1**(2006)05002.
[17] Oded Arnon, Optical Engineering, 78a Shenkin St., Tel Aviv 65223, Israel.
[18] Optimax Systems, Inc., 6367 Dean Parkway, Ontario, NY 14519-8939.
[19] P. Annis, S. Aoki, G. Brooijmans, J. Brunner, M. de Jong, J.-P. Fabre, R. Ferreira, W. Flegel, D. Frekers, G. Gregoire, M. Gruw, J. Herin, K. Hoepfner, M. Kobayashi, J. Konijn, V. Lemaitre, P. Lendermann, D. Macina, R. Meijer Drees, H . Meinhard L. Michel, C. Mommaert, K. Nakamura, M. Nakamura, T. Nakano, K. Niwa, E. Niu, J. Panman, F. Riccardi, D. Rondeshagens, 0. Sate, G. Stefanin, M. Vander Donckt, P Vilain, G. Wilquet, K. Winter, H.T. Wong.
Nucl. Instr. Meth. **A367**(1995)367.
[20] Santa Barbara Instrument Group (SBIG), Inc., 147-A Castilian Drive, Santa Barbara, CA 93117.
[21] J.F. Ziegler, SRIM 2008; http://www.srim.org.

[22] S.F. Biagi; Nucl. Instr. Meth. **A421**(1999)234; http://ref.web.cern.ch/ref/CERN/CNL/2000/001/magboltz.
[23] G. Laczko, V. Dangendorf, M. Kramer, D. Schardt, K. Tittelmeier; arxiv.org/pdf/nucl-ex/0310016.
[24] W.A. Watson *et al*., CODA; Proceedings of the Real Time 1993 Conference, p. 296; G. Heyes et al.; Proceedings of the CHEP Conference, 1994, p. 122; G. Heyes *et al*., Proceedings of the real time 1993 Conference, p. 382; D.J. Abbott *et al*.; 11th IEEE NPSS Real Time 1999 Conference, JLab-TN-99-12, 1999. http://coda.jlab.org.
[24] Ch. Sun, Ph.D. Dissertation, Duke University (2009).
[25] P. Fonte, V. Peskov, B. Ramsey; IEEE Trans. Nucl. Science, **46**(1999)312.